\documentclass[aps,prc,twocolumn,superscriptaddress,preprintnumbers,amsmath,amssymb,floatfix,showpacs,showkeys,nofootinbib]{revtex4}

\usepackage{graphicx, fink}
\usepackage{epsfig}
\usepackage{dcolumn}
\usepackage{bm}
\usepackage{longtable}
\usepackage{color}
\usepackage{float}
\usepackage{comment}
\usepackage{amsmath}
\usepackage{amsfonts}
\usepackage{xspace}
\usepackage{multirow}
\usepackage{nicefrac}
\usepackage[normalem]{ulem}

\begin{document}

\newcommand{\nuc}[2]{\ensuremath{^{#1}}#2}
\newcommand{\etal}{{\em et~al.}}
\newcommand{\geant} {{\bf {G}\texttt{\scriptsize{EANT}}4}}
\newcommand{\alphae}{\ensuremath{\alpha_{E1}}}
\newcommand{\betam}{\ensuremath{\beta_{M1}}}
\newcommand{\ChiEFT}{\ensuremath{\chi}EFT\xspace}

\preprint{MAX-lab $^{2}$H$(\gamma,\gamma)$ Summary Article for arXiv}

\title{Compton Scattering from the Deuteron below Pion-Production Threshold}

\author{\mbox{L.~S.~Myers}}
  \altaffiliation{Present address: Thomas Jefferson National Accelerator Facility, Newport News, VA 23606, USA}
\affiliation{Department of Physics, University of Illinois at Urbana-Champaign, Urbana, IL 61801, USA}
\author{\mbox{J.~R.~M.~Annand}}
\affiliation{School of Physics and Astronomy, University of Glasgow, Glasgow G12 8QQ, Scotland UK}
\author{\mbox{J.~Brudvik}}
\affiliation{MAX IV Laboratory, Lund University, SE-221 00 Lund, Sweden}
\author{\mbox{G.~Feldman}}
\affiliation{Institute for Nuclear Studies, Department of Physics, The George Washington University, Washington DC 20052, USA}
\author{\mbox{K.~G.~Fissum}}
  \altaffiliation{Corresponding author; \texttt{kevin.fissum@nuclear.lu.se}}
\affiliation{Department of Physics, Lund University, SE-221 00 Lund, Sweden}
 \author{\mbox{H.~W.~Grie{\ss}hammer}}
\affiliation{Institute for Nuclear Studies, Department of Physics, The George Washington University, Washington DC 20052, USA}
\author{\mbox{K.~Hansen}}
\affiliation{MAX IV Laboratory, Lund University, SE-221 00 Lund, Sweden}
\author{\mbox{S.~S.~Henshaw}}
  \altaffiliation{Present address: National Security Technologies, Andrews AFB, MD 20762, USA}
\affiliation{Department of Physics, Duke University, Durham, NC 27708, USA}
\author{\mbox{L.~Isaksson}}
\affiliation{MAX IV Laboratory, Lund University, SE-221 00 Lund, Sweden}
\author{\mbox{R.~Jebali}}
\affiliation{School of Physics and Astronomy, University of Glasgow, Glasgow G12 8QQ, Scotland UK}
\author{\mbox{M.~A.~Kovash}}
\affiliation{Department of Physics and Astronomy, University of Kentucky, Lexington, KY 40506, USA}
\author{\mbox{M.~Lundin}}
\affiliation{MAX IV Laboratory, Lund University, SE-221 00 Lund, Sweden}
\author{\mbox{D.~G.~Middleton}}
\affiliation{Kepler Centre for Astro and Particle Physics, Physikalisches Institut, Universit{\"a}t T{\"u}bingen, D-72076 T{\"u}bingen, Germany}
\author{\mbox{A.~M.~Nathan}}
\affiliation{Department of Physics, University of Illinois at Urbana-Champaign, Urbana, IL 61801, USA}
\author{\mbox{B.~Schr\"oder}}
  \affiliation{MAX IV Laboratory, Lund University, SE-221 00 Lund, Sweden}
\affiliation{Department of Physics, Lund University, SE-221 00 Lund, Sweden}
\author{\mbox{S.~C.~Stave}}
  \altaffiliation{Present address: Pacific Northwest National Laboratory, Richland, WA 99352, USA}
\affiliation{Department of Physics, Duke University, Durham, NC 27708, USA}

\collaboration{The COMPTON@MAX-lab Collaboration}
\noaffiliation

\date{\today}

\begin{abstract}

Differential cross sections for elastic scattering of photons from the deuteron 
have recently been measured at the Tagged-Photon Facility at the MAX IV 
Laboratory in Lund, Sweden. These first new measurements in more than a decade 
further constrain the isoscalar electromagnetic polarizabilities of the nucleon and provide the 
first-ever results above 100~MeV, where the sensitivity to the polarizabilities 
is increased. We add 23 points between 70 and 112~MeV, at angles $60^\circ$,
$120^\circ$ and $150^\circ$. Analysis of these data using a Chiral Effective Field Theory 
indicates that the cross sections are both self-consistent and consistent with
previous measurements. Extracted values of $\alpha_{s}$ = 
[12.1 $\pm$ 0.8(stat) $\pm$ 0.2(BSR) $\pm$ 0.8(th)] $\times$ 10$^{-4}$ fm$^3$ 
and $\beta_{s}$ = [2.4 $\pm$ 0.8(stat) $\pm$ 0.2(BSR) $\pm$ 0.8(th)] $\times$ 
10$^{-4}$ fm$^3$ are obtained from a fit to these 23 new data points. This paper presents in 
detail the experimental conditions and the data analysis used to extract 
the cross sections.


\end{abstract}


\maketitle

\section{\label{section:Introduction}Introduction}

A new result on the extraction of the nucleon electromagnetic polarizabilities
was recently reported, based on recent measurements of Compton scattering from the 
deuteron~\cite{myers2014a}. This paper presents in detail the motivation, configuration, data analysis and
critical evaluation of the experiment reported therein, as well as a parameter
extraction using only the new data.

Low-energy nuclear Compton scattering $\gamma X\to\gamma X$ explores how the internal
degrees of freedom of the target behave in the electric and magnetic fields of
a real external photon. Since these fields induce radiation multipoles by
displacing the target constituents, the energy dependence of the emitted
radiation provides a stringent test of the symmetries and strengths which govern
the interactions of the constituents with each other and with the photon; see
e.g.~a recent review~\cite{griesshammer2012}.  

The proton response can be measured directly and cleanly using a
${}^1\mathrm{H}$ target. The neutron, however, is much more difficult to study
because it is unstable outside the nucleus and its coupling to photons is much 
weaker. Embedding the neutron into a stable nucleus allows 
its two-photon response to be reconstructed. An added benefit of this approach is that the signal from 
the neutron is enhanced through its interference with the contributions from 
the charged proton. It also enables one to probe how the photons couple to the 
charged pion-exchange currents which provide the bulk of nuclear binding. The deuteron, which is 
the simplest stable few-nucleon system, is an ideal target for Compton-scattering 
experiments as it provides a conceptually clean probe of our understanding of both 
single-hadron and few-nucleon physics at low energies.

After subtracting binding effects, theorists utilize such data to extract the
two-photon response of the individual nucleon to the static fields;
see~\cite{griesshammer2012, Hildebrandt:2003fm} for details. 
First, one subtracts the Powell amplitudes~\cite{powell1949} for photon 
scattering on a point-like spin-$\nicefrac{1}{2}$ nucleon with an anomalous magnetic moment. 
The remainder is then expanded into energy-dependent radiation
multipoles of the incident and outgoing photon fields. Finally, these
coefficients are extrapolated to the values at zero photon frequency
$\omega$. In that limit, the leading contributions are
quadratic in $\omega$. Their coefficients are called the static
electric dipole polarizability $\alphae$ and the static magnetic
dipole polarizability $\betam$ and can be separated by different angular
dependences.
They parametrize the stiffness of the nucleon against $E1\to E1$ and
$M1\to M1$ transitions at zero photon energy, respectively. 

A host of information about the hadron response is thus compressed into
$\alphae$ and $\betam$, often referred to as ``the polarizabilities''.
They are experimentally not directly
accessible since assumptions about the energy dependence and conventions on how
to separate one- and two-photon physics enter. Nonetheless, they summarize
information on the entire spectrum of nucleonic excitations. By comparing the
quantities extracted from data with fully dynamical lattice QCD extractions
which are anticipated in the near future~\cite{lujan2014,Lujan:2014qga,detmold2012}, 
the polarizabilities will offer a stringent test of our understanding of Quantum Chromodynamics
(QCD). Most notable is the
opportunity to explore the two degrees of freedom with the lowest excitation
energy, namely the pion cloud around the nucleon and the $\Delta(1232)$
excitation. Since both of these are dominated by isospin-symmetric
interactions, differences between proton and neutron polarizabilities signal
the breaking of isospin and chiral symmetry, in concert with such effects from short-distance physics.
Besides being fundamental nucleon properties, $\alphae$ and $\betam$ also play
a role in theoretical studies of the Lamb shift of muonic hydrogen and of the
proton-neutron mass difference, and dominate the uncertainties of
both~\cite{Pachucki:1999,Carlson:2011dz,Birse:2012eb,WalkerLoud:2012bg}.

As recently reviewed in Ref.~\cite{griesshammer2012}, a statistically
consistent proton Compton-scattering database contains a cornucopia of points
between $30$ and $170$~MeV, with good angular coverage and statistical uncertainties
usually around $5$\%. Based on this extensive set, McGovern et al.~\cite{mcgovern2012}
extracted the proton polarizabilities in 
Chiral Effective Field Theory (\ChiEFT), the extension of Chiral Perturbation
Theory to include baryons, as \footnote{We use
  the canonical units of $10^{-4}$~fm$^3$ for the nucleon polarizabilities
  throughout.}
\begin{equation}
  \label{eq:AlphaBetaP}
  \begin{split}
    &\alpha_p = 10.65 \pm 0.35(\text{stat}) \pm 0.2(\text{BSR}) \pm
    0.3(\text{th}) \\ 
    &\beta_p = 3.15 \mp 0.35(\text{stat}) \pm 0.2(\text{BSR}) \mp
    0.3(\text{th}),
  \end{split}
\end{equation}
with $\chi^2=113.2$ for $135$ degrees of freedom. Theoretical uncertainties
are separated from those induced by application of the Baldin Sum Rule (BSR) for the
proton~\cite{baldin1960}. The BSR is a variant of the optical theorem which uses proton
photoabsorption cross-section data to provide the constraint that $\alpha_p +
\beta_p=13.8 \pm 0.4$~\cite{deleon2001}.

By contrast, the neutron polarizabilities, as extracted from deuteron Compton scattering, are poorly determined.
The calculations related to neutron polarizabilities appear to be
theoretically well under control~\cite{griesshammer2012}, but 
the experimental deuteron data are of smaller quantity and poorer 
quality than those of the proton.
Three experiments have thus far constituted
the world data: the pioneering effort of
Lucas et al.~at $49$ and $69$~MeV~\cite{lucas1994}; the follow-up measurement
by Lundin et al.~\cite{lundin2003} which covered similar energies
and angles; and the extension to $95$~MeV by Hornidge et
al.~\cite{hornidge2000}. This statistically consistent database contains
only $29$ data points at $4$ energies between $49$ and $95$~MeV,
with limited angular coverage, typical statistical uncertainties of more than
$7$\%, and typical systematic uncertainties in excess of $4$\%. From these data, the
isoscalar (average) nucleon polarizabilities were extracted using the same
\ChiEFT methodology as for the proton as
\begin{equation}
  \label{eq:AlphaBetaS1}
  \begin{split}
    &\alpha_s = 10.9 \pm 0.9(\text{stat}) \pm 0.2(\text{BSR}) \pm
    0.8(\text{th}) \\ 
    &\beta_s = 3.6 \mp 0.9(\text{stat}) \pm 0.2(\text{BSR}) \mp
    0.8(\text{th}),
  \end{split}
\end{equation}
with $\chi^2=24.2$ for $25$ degrees of freedom~\cite{griesshammer2012}. 
The result is again constrained by a BSR for the nucleon. The
isoscalar value
\begin{equation}
  \label{eq:BSRS}
 \alpha_s + \beta_s=14.5 \pm 0.4
\end{equation}
is found by combining the proton BSR above with empirical
partial-wave amplitudes for pion photoproduction on the neutron which lead to
the neutron sum rule $\alpha_n + \beta_n=15.2 \pm
0.4$~\cite{levchuk2000}. The uncertainty in the neutron BSR is highly correlated with that for the
proton.

Combining isoscalar and proton polarizabilities, Eqs.~\eqref{eq:AlphaBetaP} and
\eqref{eq:AlphaBetaS1}, leads to neutron values
\begin{equation}
  \label{eq:AlphaBetaN}
  \begin{split}
    &\alpha_n = 11.1 \pm 1.8(\text{stat}) \pm 0.2(\text{BSR}) \pm
    0.8(\text{th}) \\ 
    &\beta_n = 4.1 \mp 1.8(\text{stat}) \pm 0.2(\text{BSR}) \mp 0.8(\text{th}),
  \end{split}
\end{equation}
which are clearly dominated by the statistical uncertainties, which in turn are much larger than for the proton; see Eq.~\eqref{eq:AlphaBetaP}. 

An alternative extraction of neutron polarizabilities from the 7 data points 
measured in quasi-elastic \nuc{2}{H}($\gamma$,$\gamma^\prime$$n$)$p$ above 200 MeV~\cite{Kossert:2002ws} 
is consistent with these numbers.  Again using
the neutron BSR constraint, one finds:
\begin{equation}
  \label{eq:AlphaBetaKossert}
  \begin{split}
    &\alpha_n = 12.5 \pm 1.8(\text{stat}){}^{+1.1}_{-0.6}(\text{sys})\pm
    1.1(\text{th}) \\ 
    &\beta_n = 2.7 \mp 1.8(\text{stat}){}_{-1.1}^{+0.6}(\text{sys})\mp
    1.1(\text{th}),
  \end{split}
\end{equation}
where the theory uncertainty may be underestimated~\cite{Lvov}. 
No extractions from heavier nuclei exist; good Compton data is available on $^6$Li~\cite{Myers:2012xw,Myers:2014qaa}, but comparable data have not been published for any other few-nucleon systems.
A third technique, extracting the neutron polarizabilities from scattering 
from the Coulomb field of heavy nuclei, appears to be plagued by poorly understood
systematic effects; see e.g.~\cite{griesshammer2012}. 

New deuteron data of good quality and reproducible systematic uncertainties are
therefore necessary to see commonalities and differences in the two-photon
responses of protons and neutrons. The experiment detailed in this paper effectively
doubled the deuteron Compton database and significantly reduced the uncertainties of the
neutron polarizabilities. These data overlap the previous sets at lower
energies and add points up to $112$~MeV, with statistical and systematic
uncertainties on par with preceding measurements. The extension to
higher energies is particularly important since the sensitivity of the cross
sections to the polarizabilities increases roughly with the square of the photon energy.

The resulting augmented world database is statistically consistent and
recently resulted in a new extraction of the isoscalar polarizabilities in
Refs.~\cite{myers2014a,futuretheory} as
\begin{equation} 
  \label{eq:AlphaBetaNewBaldin}
  \begin{split}
    &\alpha_s = 11.1 \pm 0.6(\text{stat}) \pm 0.2(\text{BSR}) \pm
    0.8(\text{th}) \\
    &\beta_s = 3.4 \mp 0.6(\text{stat}) \pm 0.2(\text{BSR}) \mp
    0.8(\text{th}).
  \end{split}
\end{equation}
Our new measurement thus reduces the statistical uncertainties in $\alpha_s$ and $\beta_s$ by a factor of
$1/3$.  For the very first time, the uncertainty is now dominated by the theoretical
uncertainties of the extraction.

While some aspects of our findings have been summarized briefly in a recent
publication~\cite{myers2014a}, we now provide a more detailed description of the entire experimental 
effort, including complementary information to aid in the interpretation of the results.
In Sections~\ref{section:Experiment} and
\ref{section:data_analysis}, we present the experimental setup at MAX IV and
the data-analysis process, and pay special attention to yield corrections and
systematic uncertainties. Section~\ref{section:results} contains the resulting
cross sections and an extraction of the polarizabilities, 
focusing on the self-consistency of this data set and its agreement with previous measurements.

\section{\label{section:Experiment}Experiment}

\begin{figure}
\begin{center}
\includegraphics[width=0.4\textwidth]{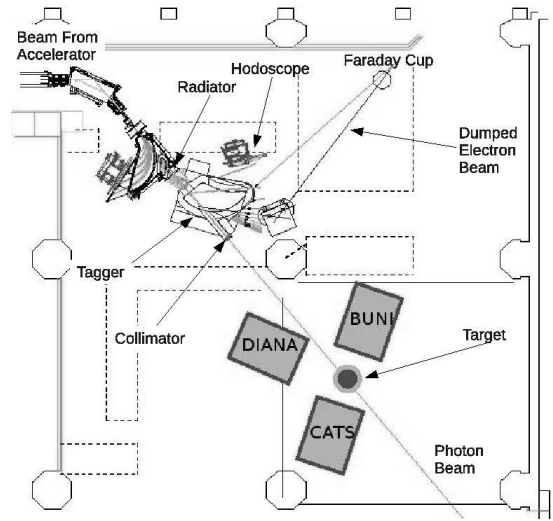}
\caption{
The layout of the experimental area showing the location of the tagging 
spectrometer, focal-plane hodoscope, deuterium target, and NaI(Tl) detectors 
labeled DIANA, BUNI, and CATS.
\label{figure:figure_01_setup}}
\end{center}
\end{figure}

The experiment was performed at the Tagged-Photon Facility~\cite{adler2012}
located at the MAX IV Laboratory \cite{eriksson2014} in Lund, Sweden.  A
pulse-stretched electron beam~\cite{lindgren2002} with 
a typical current of 15~nA and a duty factor of 45\% was used to produce 
quasi-monoenergetic photons via the 
bremsstrahlung-tagging technique~\cite{adler1990,adler1997}. 
The basic parameters of the electron and resulting tagged-photon 
beams are given in Table~\ref{table:beam_parameters} for the first and 
second run periods, RP1 and RP2.
An overview of the 
experimental layout is shown in Fig.~\ref{figure:figure_01_setup}.

\begin{table}
\caption{\label{table:beam_parameters}
Basic parameters of the electron beam and the tagged-photon beam for the
two run periods, RP1 and RP2.}
\begin{ruledtabular}
\begin{tabular}{lrr}
                            &     RP1    &     RP2 \\
\hline
   $E_{\rm electron}$ [MeV]  &              144    &              165 \\
        $E_{\gamma}$ [MeV]  &         65 -- 97    &        81 -- 115 \\
$E_{\gamma\rm, bin}$ [MeV]  &        69.6, 77.8   &       85.8, 94.8 \\
                           &        86.1, 93.7   &     103.8, 112.1 \\
$\Delta E_{\gamma\rm, bin}$ [MeV]  &       $\sim$8.0    &   $\sim$8.5 \\
\end{tabular}
\end{ruledtabular}
\end{table}

\begin{figure}
\begin{center}
\includegraphics[width=0.48\textwidth]{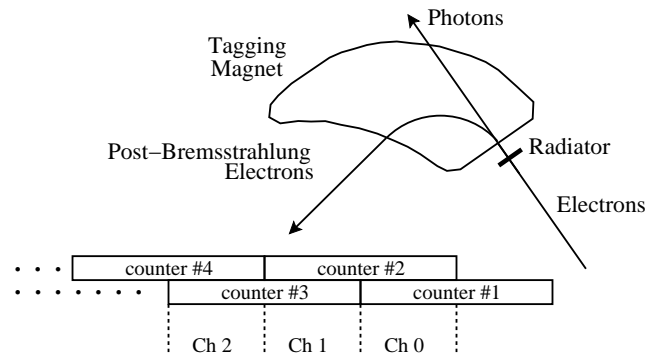}
\caption{
Enlarged diagram of the tagger magnet and FP hodoscope portion of the experimental layout (not to scale).
\label{figure:figure_02_tagger}}
\end{center}
\end{figure}

The tagging magnet and focal-plane (FP) hodoscope \cite{vogt1993}
were used extensively at the Saskatchewan Accelerator Laboratory prior to their
use at the MAX IV Laboratory. The dipole field of the magnet is used to momentum analyze the
post-bremsstrahlung electrons, which are detected in the FP
hodoscope by 63 plastic scintillators. The scintillators are 25 mm wide and 3 mm
thick and arranged into two rows. The rows are offset by 50\% of the scintillator
width, with each overlap defining a FP channel (see Fig.~\ref{figure:figure_02_tagger}). The typical
width of a FP channel was $\sim$400~keV and the nominal electron
rate was 1~MHz/channel. As the Compton counting rate is low, the focal plane was 
subdivided into four bins. 
The central photon energies for each bin, as well as the average bin width, are
given in Table~\ref{table:beam_parameters}.

The size of the photon beam was defined by a tapered tungsten-alloy primary 
collimator of 19~mm nominal diameter.  The primary collimator was followed 
by a dipole magnet and a post-collimator which were used to remove any charged 
particles produced in the primary collimator. The beam spot at the target 
location was approximately 60~mm in diameter.

The tagging efficiency~\cite{adler1997} is the ratio of the number of tagged
photons which struck the target to the number of post-bremsstrahlung electrons 
which were registered by the associated FP channel. It 
depends on the collimator size and the electron-beam energy. It was 
measured during the experiment start-up with the three large-volume NaI(Tl) 
photon spectrometers placed directly in the low-intensity photon beam and was 
monitored during data collection on a daily basis via dedicated 
measurements with a compact lead-glass photon detector, which was easily raised 
into and lowered out of the photon beam. 

The liquid deuterium target used in this experiment was based on a design 
used in previous measurements~\cite{lundin2003}, but with measures developed to
eliminate ice build-up on the target endcaps. These measures included a
several-day bake-out of the vacuum vessel to reduce internal 
gases, thicker Kapton foils for the vacuum chamber windows, and the implementation of a N$_2$
gas shield. These last two measures reduce the penetration of water vapor from the 
air into the insulation vacuum surrounding the cell. The cell was cylindrical,
150~mm long and 68~mm in diameter. The spherical endcaps were convex so that
the total length of the cell was 170~mm. The cell was oriented so that its central axis was collinear with the beam line.
The housing chamber was constructed of 
stainless steel with a thickness of $\sim$1~mm in the vicinity of the scattering
plane and $\sim$2~mm elsewhere.

Three large-volume, segmented NaI(Tl) detectors labeled BUNI~\cite{miller1988}, 
CATS~\cite{wissmann1994}, and DIANA~\cite{myers2010} in 
Fig.~\ref{figure:figure_01_setup} were used to detect the Compton-scattered 
photons. The detectors were located at laboratory angles of 60$^\circ$, 
120$^\circ$, and 150$^\circ$.
These detectors were each composed of a NaI(Tl) core  
surrounded by optically isolated, annular NaI(Tl) segments.
The cores of the BUNI and CATS detectors each measure
26.7~cm in diameter, while the core of the DIANA detector measures 48.0~cm.
The depth of all three detectors is greater than 20 radiation lengths.
The annular segments are 11~cm thick on the BUNI 
and CATS detectors and 4~cm thick
on the DIANA detector. Additionally, BUNI and CATS each has a plastic-scintillator annulus
that surrounds the NaI(Tl) annulus.
Each detector was shielded by lead with a front aperture that defined
the detector acceptance. A plastic-scintillator paddle was placed in front of the 
aperture to identify and veto charged particles. The detectors have an energy 
resolution of better than 2\% at energies near 100 MeV. Such resolution 
is necessary to separate unambiguously the elastically scattered photons from 
those originating from the breakup of deuterium.

The signals from each detector were passed to analog-to-digital converters (ADCs) 
and time-to-digital converters (TDCs) and the data were recorded on an event-by-event 
basis.
The experimental data were collected in two separate four-week run
periods. The first run period employed an electron-beam energy of 144~MeV; the
second run period used 165~MeV.
The first week of each period was dedicated to in-beam studies (see below), 
measurements of $^{12}$C($\gamma$,$\gamma$) \cite{myers2014} to establish the absolute
normalization and systematics of the setup, and cooling of the liquid-deuterium target. 
The remaining three weeks were used to perform the measurements on deuterium reported in this 
article.

\section{\label{section:data_analysis}Data Analysis}

The Compton scattering cross section can be written as
\begin{equation}
  \label{equation:Compton_scattering_cross_section}
  \frac{d\sigma}{d\Omega} = (\frac{Y}{\Omega_{\rm eff}}) \frac{1}{N_\gamma} \frac{1}{\kappa_{\rm eff}} f_{\rm R} f_{\rm T},
\end{equation}
where $(Y/\Omega_{\rm eff})$ is the scattered-photon yield normalized to the 
effective solid-angle acceptance of the detector, $N_\gamma$ is the number 
of tagged photons incident upon the target, and $\kappa_{\rm eff}$ is the 
effective thickness of the target (the number of nuclei per unit area).
$f_{\rm R}$ and $f_{\rm T}$ represent correction factors due to rate- and
target-dependent effects, respectively, as explained below.

\subsection{\label{subsection:yield_extraction}Acceptance-normalized yield}

\subsubsection{\label{subsubsection:scattered_photon_yield}Scattered-photon yield}

\begin{figure}
\begin{center}
\resizebox{0.5\textwidth}{!}{\includegraphics{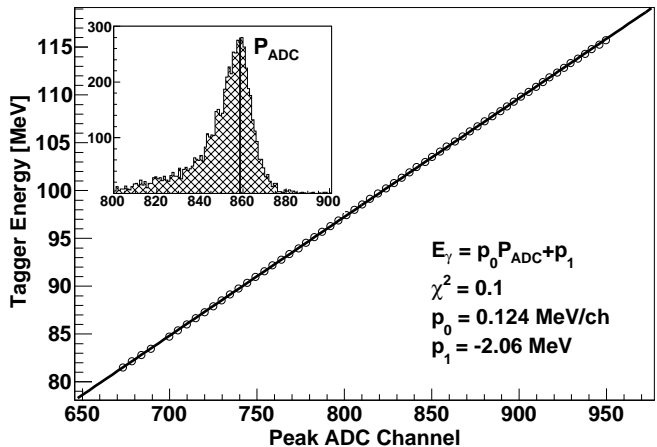}}
\caption{
\textcolor{black}{A plot of the predicted photon energy for each FP channel
versus the peak position of the ADC spectrum (P$_{\rm ADC}$) for the same channel. The plot is fit with a linear
function to determine the calibration of the NaI(Tl) detector. (Inset) A typical tagged-photon
ADC spectrum for BUNI showing the location of the ADC channel corresponding to the peak (P$_{\rm ADC}$).}
\label{figure:figure_03_calib}}
\end{center}
\end{figure}

During the experiment, the ADCs allowed reconstruction of the scattered-photon 
energies, while the TDCs enabled coincident timing between the NaI(Tl) detectors 
and the FP hodoscope. The energy calibration of each of the NaI(Tl) 
detectors was determined by placing it directly into the reduced-intensity photon 
beam and observing its response as a function of tagged-photon energy. 
To calibrate each ADC, a spectrum was filled (see the inset to
Fig.~\ref{figure:figure_03_calib}) for each FP channel by selecting only 
on tagged photons for that channel. The position of the peak (in ADC channels) was
then plotted against the expected photon energy, as determined from the tagging
magnet field map, for all the FP channels. As an example, the calibration for BUNI is 
shown in Fig.~\ref{figure:figure_03_calib}. 

\begin{figure}
\begin{center}
\resizebox{0.5\textwidth}{!}{\includegraphics{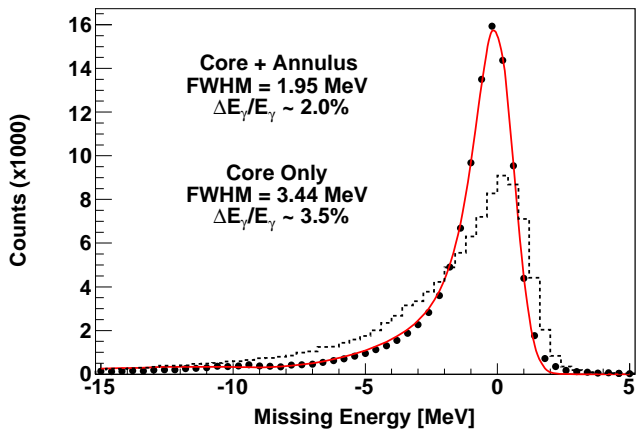}}
\caption{
\textcolor{black}{(Color online) In-beam detector response as a function of missing energy for the BUNI
detector. The data points are the result of summing the energy deposition in both the
NaI(Tl) core and annulus. The solid red curve is the simulated \geant\ detector response fitted to the 
data. The dashed line is the ME spectrum obtained by analyzing only the core
crystal. The addition of the annulus energy improves the FWHM by almost 50\%.}
\label{figure:figure_04_inbeam}}
\end{center}
\end{figure}

Missing energy (ME) was defined to be the difference between the expected energy of the detected
photon (as determined from the tagger magnet and FP hodoscope placement) and the energy deposited by 
the photon in the NaI(Tl) detector. 
In both BUNI and CATS, the energy deposition in the 
annulus was added to the core energy deposition to improve the resolution.
The measured in-beam response for the BUNI detector, together with a fitted 
\geant\ \cite{agostinelli2003} simulation of this response, is shown in 
Fig.~\ref{figure:figure_04_inbeam}. The \geant\ simulation 
output was determined for the case of the BUNI detector positioned directly in 
the photon beam. This ``intrinsic" simulation was then smeared with a Gaussian 
function according to
\begin{equation}
  \label{equation:convolution}
  \mathcal{R}_{i}^{\rm inbeam} = \sum\limits_j \frac{p_{\rm 1}}{E_j p_{\rm 3}} e^{\frac{(E_i-E_j-p_{\rm 2})^{2}}{-2(E_jp_{\rm 3})^2}} \mathcal{S}_j^{\rm inbeam},
\end{equation}
where $E_{i \left( j \right) }$ was the central energy of bin $i$($j$), 
$\mathcal{S}_j^{\rm inbeam}$ was the number of counts in bin $j$ of the simulated 
detector-response spectrum, and $p_{\rm 1,2,3}$ were fitting parameters that 
accounted for the individual characteristics of each NaI(Tl) detector, such 
as non-uniform doping of the crystal, that are difficult to model in \geant. This
process was repeated for CATS and DIANA.

It was observed that the gain of the NaI(Tl) detectors drifted 
over the course of the run periods. In order to correct for this gain drift,
cosmic-ray data were collected immediately after each in-beam calibration run
and prior to moving the detector to its scattering location. 
``Straight-through'' cosmic rays, defined by requiring a large energy 
deposition in annular segments on opposite sides of the core crystal, were 
selected because these events should have a constant energy deposition in the 
detector. The gain drift for each PMT for each run could be determined by 
monitoring the location and shape of the resulting cosmic-ray peaks. These
gain-drift corrections were then applied to the data.

The energy calibration of the tagger focal plane was confirmed by observing 
highly energetic capture photons from the $\pi^{-} + d \rightarrow \gamma + 2n$ 
reaction. This reaction channel was present as the untagged bremssstrahlung 
spectrum extended beyond pion production threshold energy, and the most probable energy of the capture photon is 
$\sim$131~MeV~\cite{gabioud1979}. The agreement between the absolute photon energy 
and that reconstructed from the tagger FP energy calibration 
was better than 1\%. 

Large backgrounds arose when the beam intensity was increased from 10-100~Hz 
per FP channel (for in-beam runs) to $\sim$1~MHz per channel (for scattering runs). 
The dominant sources of background were untagged bremsstrahlung photons (which scaled with the 
beam intensity) and cosmic rays. 
These backgrounds obscured the timing and ME peaks of
the elastic photons in the TDC and ADC spectra, respectively. Cosmic rays
deposit a large amount of energy in the detector overall and in the annular
segments in particular.
A cut placed on the NaI(Tl) annulus (BUNI and DIANA) or the  plastic-scintillator 
(CATS) annulus removed $\sim$95\% of the cosmic-ray background from 
the scattering data. An additional cut placed on the thin 
plastic-scintillator paddle in front of each of the NaI(Tl) detectors removed
charged particles. These cuts reduced the number of events by $\sim$50\%.

\begin{figure}
\begin{center}
\resizebox{0.5\textwidth}{!}{\includegraphics{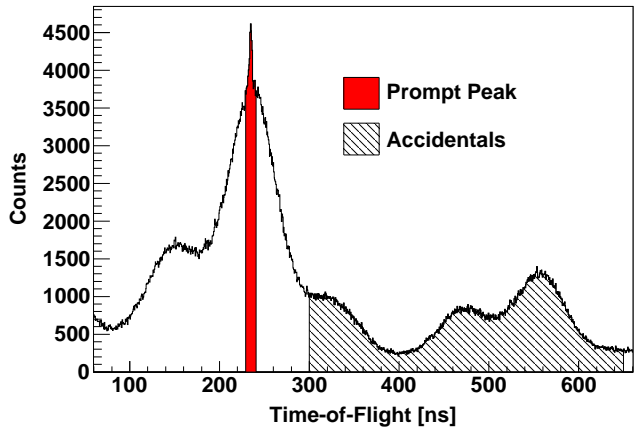}}
\caption{(Color online) The FP TDC spectrum for the scattering data.
The prompt (red) and the accidental (shaded) windows are indicated.
\label{figure:figure_05_tdc}}
\end{center}
\end{figure}

In order to further reduce the untagged bremsstrahlung background
in the FP TDC spectrum, a cut was placed on the energy deposited in the NaI(Tl) detector.
Selecting only events with an energy deposition $E_{\rm min} \le E \le E_{\rm max}$, where $E_{\rm min(max)}$
is the minimum (maximum) tagged-photon energy,
enabled the prompt peak to be identified in the FP TDC spectrum\footnote{
The structure seen in the TDC spectrum was a result
of the incomplete filling of the pulse-stretcher ring ($T$=108~ns) and the
3.3~MHz extraction shaker ($T$=305~ns)~\cite{myers2013}.} 
(see Fig.~\ref{figure:figure_05_tdc}). 
This prompt peak represented coincidences 
between post-bremsstrahlung electrons in the FP hodoscope and 
elastically scattered photons in the NaI(Tl) detectors.

For each NaI(Tl) 
detector and for each FP channel, events occurring within the 
prompt peak were selected and a prompt ME spectrum was filled. The process
was repeated for a second cut placed on a purely accidental timing region, 
and an accidental spectrum was filled. 

For each run period, data collected during dedicated, beam-off 
cosmic-ray runs were utilized to remove the $\sim$5\% of 
cosmic-ray events that survived the annulus cuts. For each NaI(Tl) 
detector and for each FP channel, these data were subjected 
to the same annulus cuts above, and cosmic-ray spectra were filled. 
These spectra were scaled by the ratio of events with an energy exceeding 
the electron-beam energy in the prompt spectra to those in the cosmic-ray 
spectra, and then subsequently subtracted from the prompt spectra.
The procedure was repeated for the accidental spectra. In this way, prompt 
and accidental spectra free from cosmic rays were produced.

\begin{figure}
\begin{center}
\resizebox{0.5\textwidth}{!}{\includegraphics{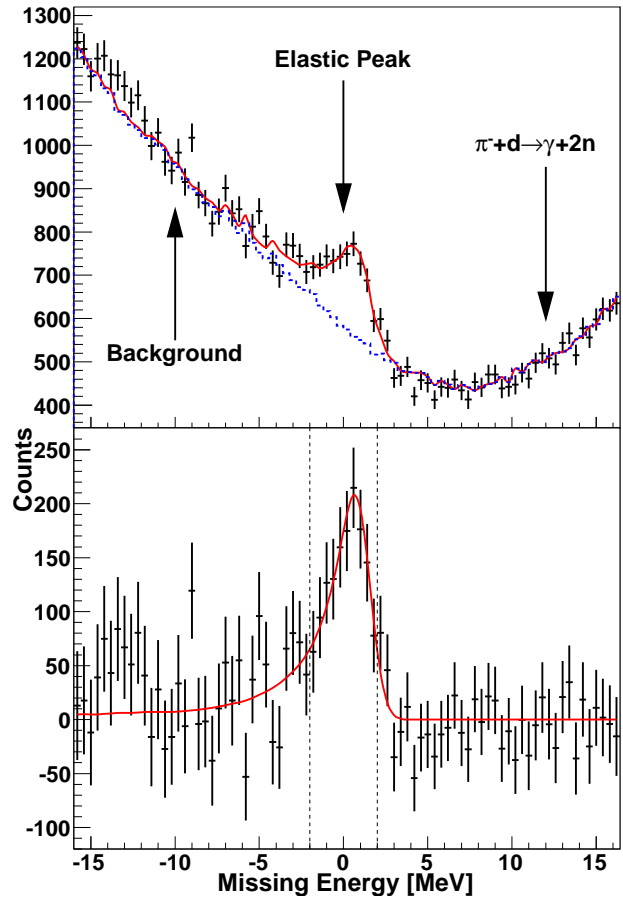}}
\caption{(Color online) Top panel: The result of fitting the sum (solid) of the 
accidental spectrum (dashed) and the fitted detector response to the 
prompt spectrum. Bottom panel: The true 
elastic peak (prompts minus accidentals) together with the fitted detector 
response (solid). The vertical, dashed lines indicate the region of integration 
over which the yield extraction occurs, $-$2.0~MeV~$\leq$~ME~$\leq$~2.0~MeV.
\label{figure:figure_06_fitting}}
\end{center}
\end{figure}

The \geant\ in-beam simulation was extended to reflect the 
experimental setup for scattering runs. A numerical function 
$\mathcal{F}_{i}$ was defined by
\begin{equation}
  \label{equation:scattering}
  \mathcal{F}_{i} = p_{\rm 0}\mathcal{A}_i + \mathcal{R}_i^{\rm scatter},
\end{equation}
where $\mathcal{A}_i$ was the number of counts in bin $i$ of the accidental 
spectrum, $p_{\rm 0}$ was the scale factor of the accidentals, and 
$\mathcal{R}_i^{\rm scatter}$ was given by Eq.~\eqref{equation:convolution} 
using the scattering response spectrum from the \geant\ simulation (see 
the top panel of Fig.~\ref{figure:figure_06_fitting}). The range of the fitting 
window varied 
from as small as \mbox{[--10,+10]}~MeV to as large as \mbox{[--20,+20]}~MeV in ME. By fitting 
several windows over this range, an estimate of the dependence of the 
extracted yield on the width of the fitting window was obtained. This 
kinematic-dependent uncertainty depended strongly on the signal-to-noise ratio in 
the prompt ME spectrum and ranged from 2\% to 11\%.

The bottom panel of Fig.~\ref{figure:figure_06_fitting} shows a typical ``true'' 
scattering spectrum (prompts minus accidentals) together with the 
corresponding fitted, convoluted response function $\mathcal{R}_i^{\rm scatter}$.
The scattering yield was extracted by integrating the true spectrum over
the region of interest (ROI) ($-$2.0~MeV~$\leq$~ME~$\leq$~2.0~MeV) indicated by 
the vertical dashed lines. The ROI was not allowed to extend below $-$2.0~MeV
as photons from the photodissociation of the deuteron are kinematically allowed
in this region.  The elastically scattered photon yields determined according 
to this procedure are given in Table~\ref{table:yields_and_solid_angles}.

The $-$2.0~MeV~$\leq$~ME~$\leq$~2.0~MeV ROI was carefully
chosen, taking into account the detector resolution (FWHM \textless 2\% at 100~MeV),
to minimize the contribution of photons from $d$($\gamma$,$\gamma^\prime$)$n$$p$
to the extracted yield. Contamination from non-elastic photons was investigated
with a direct and an indirect search.
The direct method involved simulating the photons from the breakup reaction 
and adding this new lineshape to the fitting algorithm. This new lineshape was
displaced 2.2~MeV to the low-energy side of the elastic peak to account for the
reaction threshold. The data were re-fit and the contribution of the non-elastic
photons to the extracted yield was calculated. This contribution was found to 
be consistent with zero within uncertainties.
The indirect method relied on the quality of the fit shown in Fig.~\ref{figure:figure_06_fitting},
which has a typical reduced $\chi^2$ value of $\lesssim$ 1. If there were a sizable contribution
arising from a non-elastic reaction, it would cause the extracted cross sections 
to vary depending on the ROI width. Since non-elastic photons lie to the left
of the elastic peak, the lower edge of the ROI was varied by $\pm$400~keV. It was
found that the extracted cross sections with the varied ROIs agreed with the ones 
listed in this paper within uncertainties.
Thus, it was concluded that the cross sections presented here do not suffer 
from contributions from $d$($\gamma$,$\gamma^\prime$)$n$$p$.

\subsubsection{\label{subsubsection:detector_acceptance}Detector Acceptance}

The detector acceptances depended on the cuts employed during the data analysis
and the width and location of the integration ROI. The geometrical solid angle 
subtended by the NaI(Tl) detectors was corrected for the geometry of the 
experimental setup. Both of these effects were studied using the \geant\ Monte 
Carlo simulations. The effective solid angle $\Omega_{\rm eff}$ was given by
\begin{equation}
  \label{equation:effective_solid_angle}
  \Omega_{\rm eff} = \frac{N_{\rm events}^{\rm ROI}}{N_{\rm events}}~4\pi,
\end{equation}
where $N_{\rm events}$ was the total number of simulated events, and
$N_{\rm events}^{\rm ROI}$ was the number of simulated events that eventually
populated the ROI.

The cosmic-rejection cuts removed a very small number of good Compton-scattered 
photons from the ROI. 
For the CATS (60$^\circ$) detector, the plastic-scintillator annulus was 
used as a cosmic-ray veto. For the DIANA (150$^\circ$) detector, the thin NaI(Tl) annulus was used
to reject cosmic rays. In each case, the cosmic-ray veto was more than 20~cm from the 
central cylindrical symmetry axis of the detector and only $\sim$1\% of all elastically
scattered photons were rejected.
For the BUNI (120$^\circ$) detector, annular NaI(Tl) segments $\sim$13~cm from
the cylindrical symmetry axis of the detector were used for the rejection
of cosmic rays. As a result, 6\% of all scattered photons were rejected. 
However, the NaI(Tl) segments in BUNI provided sufficient energy resolution
to determine the cut placement accurately resulting in a systematic 
uncertainty of $\sim$2\%. The charged-particle veto removed $\sim$1\%
of all scattered photons.

The effective solid angle for each data point is given in 
Table~\ref{table:yields_and_solid_angles}. A sufficient number of events 
were simulated so that the statistical uncertainties
are $\leq$1\%. The systematic uncertainties include the effects of the 
cosmic-rejection cuts as well as uncertainty from measurements of the 
target-detector distance and detector-aperture diameter, 
typically $\pm$2~mm.

\begin{table*}
\caption{\label{table:yields_and_solid_angles}
Extracted yields and effective solid angle at each energy and angle. For the
yields, the first uncertainty is statistical and the second is a kinematic-dependent
systematic. For the effective solid angles, the first uncertainty is statistical
and the second is an angle-dependent systematic. The upper four
energy bins are from RP1 and the lower four from RP2 (also in Tables~\ref{table:beam_photons}, \ref{table:rate_and_target_corrs}, and \ref{table:data}).
}
\begin{ruledtabular}
\begin{tabular}{r|rr|rr|rr}
$E_{\gamma}$ &                             \multicolumn{2}{c|}{60$^\circ$} &                     \multicolumn{2}{c|}{120$^\circ$} &                      \multicolumn{2}{c}{150$^\circ$} \\
       (MeV) &                         \multicolumn{1}{c}{$Y$} &     \multicolumn{1}{c|}{$\Omega_{\rm eff}$ (msr)} &                      \multicolumn{1}{c}{$Y$} & \multicolumn{1}{c|}{$\Omega_{\rm eff}$ (msr)} &                     \multicolumn{1}{c}{$Y$} & \multicolumn{1}{c}{$\Omega_{\rm eff}$ (msr)} \\
\hline
        69.6 & 1080 $\pm$ 182 $\pm$     61 &     29.0 $\pm$ 0.3 $\pm$ 1.2 & 1106 $\pm$ 191 $\pm$  42 & 42.9 $\pm$ 0.4 $\pm$ 1.9 &                         &                          \\
        77.9 &  995 $\pm$ 137 $\pm$     59 &     25.3 $\pm$ 0.3 $\pm$ 1.1 & 1528 $\pm$ 152 $\pm$  32 & 42.9 $\pm$ 0.4 $\pm$ 1.9 & 1034 $\pm$ 142 $\pm$ 84 & 22.8 $\pm$ 0.2 $\pm$ 0.7 \\
        86.1 &  809 $\pm$  93 $\pm$     18 &     28.1 $\pm$ 0.3 $\pm$ 1.2 & 1312 $\pm$ 115 $\pm$  26 & 38.1 $\pm$ 0.4 $\pm$ 1.6 &  790 $\pm$ 115 $\pm$ 41 & 22.6 $\pm$ 0.2 $\pm$ 0.7 \\
        93.4 &  440 $\pm$  63 $\pm$     13 &     26.5 $\pm$ 0.3 $\pm$ 1.1 & 1148 $\pm$  90 $\pm$  23 & 38.2 $\pm$ 0.4 $\pm$ 1.6 &  573 $\pm$  90 $\pm$ 63 & 22.2 $\pm$ 0.2 $\pm$ 0.7 \\
\hline
        85.8 & 1669 $\pm$ 199 $\pm$    187 &     24.0 $\pm$ 0.2 $\pm$ 1.0 & 2199 $\pm$ 156 $\pm$ 187 & 41.6 $\pm$ 0.4 $\pm$ 1.8 & 1616 $\pm$ 198 $\pm$ 50 & 20.2 $\pm$ 0.2 $\pm$ 0.6 \\
        94.8 & 1639 $\pm$ 161 $\pm$    127 &     18.9 $\pm$ 0.2 $\pm$ 0.8 & 2633 $\pm$ 142 $\pm$  94 & 41.1 $\pm$ 0.4 $\pm$ 1.8 & 1587 $\pm$ 174 $\pm$ 48 & 19.8 $\pm$ 0.2 $\pm$ 0.6 \\
       103.8 & 1266 $\pm$ 117 $\pm$     30 &     21.0 $\pm$ 0.2 $\pm$ 0.9 & 1919 $\pm$ 117 $\pm$  55 & 39.2 $\pm$ 0.4 $\pm$ 1.7 & 1424 $\pm$ 141 $\pm$ 71 & 19.3 $\pm$ 0.2 $\pm$ 0.6 \\
       112.1 &  842 $\pm$  91 $\pm$     21 &     21.5 $\pm$ 0.2 $\pm$ 0.9 & 1370 $\pm$  95 $\pm$  29 & 37.7 $\pm$ 0.4 $\pm$ 1.6 & 1034 $\pm$ 115 $\pm$ 29 & 19.4 $\pm$ 0.2 $\pm$ 0.6 \\
\end{tabular}
\end{ruledtabular}
\end{table*}

\subsection{\label{subsection:scale_normalization}Scale normalization}

\subsubsection{\label{subsubsection:beam_photons}Number of beam photons}

The number of beam photons incident on the target $N_{\gamma}$ was determined 
from 

\begin{equation}
  N_{\gamma} = N_{e} \cdot \varepsilon_{\rm tag},
  \label{equation:tagging_efficiency}
\end{equation}

\noindent where $N_{e}$ is the number of post-bremsstrahlung electrons detected in each FP 
channel and $\varepsilon_{\rm tag}$ is the tagging efficiency.

\begin{figure}
  \begin{center}
  \includegraphics[trim=0cm 0cm 5cm 0cm, clip=true, width=0.5\textwidth]{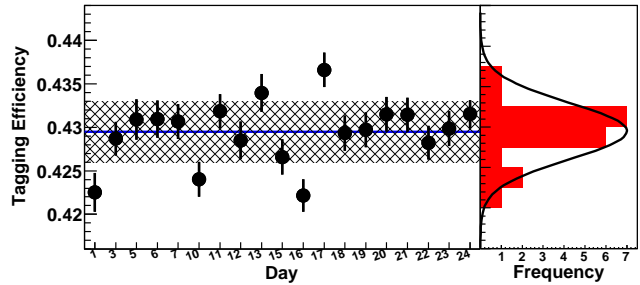}
  \caption{
  Time evolution of the tagging efficiency measured using the compact lead-glass 
  detector for the $E_{\gamma}$ = 93.4 MeV bin from RP1. The shaded region indicates
  the uncertainty due to daily variations in $\varepsilon_{\rm tag}$.
  \label{figure:figure_pbglass}}
  \end{center}
\end{figure}

The number of electrons striking the focal plane ($N_{e}$) was counted 
by the FP scalers (see Table~\ref{table:beam_photons}). The background 
rate in the focal plane, obtained from beam-off runs, was on the order of 
1~Hz per channel and was thus negligible compared to the beam-on electron rate of 
$\sim$1~MHz per channel.
The tagging efficiency $\varepsilon_{\rm tag}$ was determined from the ratio of 
the number of photons tagged by a FP channel and recorded in the in-beam 
photon detector to the number of post-bremsstrahlung electrons recorded by the same 
channel. Livetime-corrected beam-on and beam-off backgrounds ($\sim$5\%) were 
removed from the data. The focal plane was divided into four bins, each 16 channels 
wide. The tagging efficiency for each bin was taken to be the electron-weighted 
average of the tagging efficiencies for each of the 16 channels. A plot of the
daily tagging efficiency measured using the compact lead-glass detector for the 
$E_{\gamma}$ = 93.4 MeV bin is shown in Fig.~\ref{figure:figure_pbglass}.

In order to observe any systematic difference between the tagging efficiency 
determined by the NaI(Tl) detectors and the lead-glass detector, measurements
were taken with the CATS detector and immediately thereafter with the lead-glass
detector.
These data allowed for a \textless 2\% correction to be made to the compact lead-glass 
detector results, mainly due to the larger volume of the NaI(Tl) and the presence 
of the thin paddle used to veto charged particles. This correction was applied 
to the average value of the tagging efficiency determined with the compact 
lead-glass detector. One-half of the correction was assigned as a systematic 
uncertainty.

The number of post-bremsstrahlung electrons $N_{e}$, tagging efficiencies 
$\varepsilon_{\rm tag}$, and number of beam photons $N_{\gamma}$ are presented 
in Table~\ref{table:beam_photons}.

\begin{table*}
\caption{\label{table:beam_photons}
Number of post-bremsstrahlung electrons, tagging efficiencies, and number of tagged photons. 
The relative statistical and systematic uncertainties of $N_{\gamma}$ are the same
as those for $\varepsilon_{\rm tag}$. The upper four
energy bins are from RP1 and the lower four from RP2.
}
\begin{ruledtabular}
\begin{tabular}{rr|rrr|r}
$E_{\gamma}$ &           $N_{e}$ ($\times$ 10$^{13}$) & $\varepsilon_{\rm tag}$ & $\delta\varepsilon_{\rm tag}$ & $\delta\varepsilon_{\rm tag}$ &            $N_{\gamma}$ ($\times$ 10$^{12}$) \\
       (MeV) & [60$^\circ$, 120$^\circ$, 150$^\circ$] &                        &                 (statistical) &                  (systematic) & [60$^\circ$, 120$^\circ$, 150$^\circ$] \\
\hline
              69.6 &                             1.06, 1.06, 1.05 &              0.421 &                                      0.002 &               0.005 &       4.47, 4.47, 4.43 \\
              77.9 &                             1.12, 1.12, 1.11 &              0.423 &                                      0.002 &               0.005 &       4.73, 4.73, 4.69 \\
              86.1 &                          0.941, 0.941, 0.931 &              0.426 &                                      0.002 &               0.005 &       4.01, 4.01, 3.97 \\
              93.4 &                          0.782, 0.782, 0.775 &              0.425 &                                      0.002 &               0.005 &       3.32, 3.32, 3.29 \\
\hline
              85.8 &                             1.72, 1.72, 1.71 &               0.456 &                                      0.004 &              0.007 &       7.83, 7.83, 7.79 \\
              94.8 &                             1.85, 1.85, 1.84 &               0.459 &                                      0.004 &              0.008 &       8.50, 8.50, 8.45 \\
             103.8 &                             1.58, 1.58, 1.57 &               0.460 &                                      0.004 &              0.008 &       7.26, 7.26, 7.22 \\
             112.1 &                             1.37, 1.37, 1.36 &               0.458 &                                      0.005 &              0.008 &       6.24, 6.24, 6.21 \\
\end{tabular}
\end{ruledtabular}
\end{table*}

\subsubsection{\label{subsubsection:target_thickness}Effective target thickness}

The effective target thickness $\kappa_{\rm eff}$ (the number of nuclei per unit
area) was given by
\begin{equation}
\label{equation:effective_target_thickness}
\kappa_{\rm eff} = \frac{\rho~L~N_A}{A},
\end{equation}
where $\rho$ was the average density of liquid deuterium, $L$ was the effective 
target length, $N_A$ was Avogadro's number, and $A$ was the molar mass 
(4.0282~g/mol).

The target pressure and temperature were systematically recorded for  
each run. As the density of liquid deuterium is related to its 
pressure~\cite{glebe1993}, an average target density was calculated by
determining the deuterium density measured during each run and weighting this 
density by the number of FP electrons recorded in the same run. For each run period,
the average density of the liquid deuterium in the target cell was 
thus determined to be $\rho$ = (0.163 $\pm$ 0.002) g/cm$^3$. 

The cylindrical portion of the liquid deuterium target cell measured 68~mm in 
diameter and 150~mm in length. The convex endcaps each added an additional 
10~mm to the cell length at the central symmetry axis (which also corresponded 
to the photon-beam axis). Thus, the total length of the liquid deuterium target 
cell along its symmetry axis was 170~mm. Due to the cell geometry and the 
divergent nature of the photon trajectories originating from the radiator, the 
target thickness for each individual photon trajectory differed from this 
measured target length along the symmetry axis. A Monte Carlo simulation was 
thus employed to determine the effective target length for an ``average'' beam
photon. The angular distribution of the photons emanating from the radiator was 
determined from a series of tagging-efficiency measurements taken with 
photon-beam collimators of different diameters. 

The effective target length 
was determined to be $L$ = (166 $\pm$ 2)~mm. 
$\kappa_{\rm eff}$ was thus determined to be (8.10 $\pm$ 0.20) $\times$ 
10$^{23}$ (nuclei/cm$^2$), where the systematic uncertainty 
is the sum, in quadrature, of the uncertainties in $\rho$ and $L$ together with 
an additional 2\% uncertainty to account for any potential misalignment of the 
symmetry axis of the target relative to the trajectory of the photon beam. 

\subsection{\label{subsection:corrections}Corrections}

\subsubsection{\label{subsubsection:rate_dependent}Rate-dependent Corrections}

The rate-dependent correction arises from high electron rates producing dead time in the single-hit
FP TDCs and scalers.
The nominal average electron rate of 1~MHz per FP channel,
but the instantaneous rate fluctuated markedly and was as large as 4 MHz.
The origins of these 
effects has been presented in detail in Ref.~\cite{myers2013}, where a 
description of the Monte Carlo simulation of the FP electronics
used to quantify the effects is also presented. We note that this 
procedure has been used to successfully extract Compton scattering cross 
sections from carbon -- see Refs.~\cite{myers2014,preston2014}. A short summary
is presented below.

The rate-dependent correction was the product of three terms
\begin{equation}
  \label{equation:rate_dependent_correction}
  f_{\rm R} = f_{\rm ghost} f_{\rm stolen} f_{\rm missed},
\end{equation}
where $f_{\rm ghost}$ was the correction arising from accidental coincidences 
between the front and back scintillator planes in the FP hodoscope, 
$f_{\rm stolen}$ corrected for prompt electrons that are not observed in the 
prompt peak because an accidental electron stopped the single-hit TDC 
previously, and $f_{\rm missed}$ accounted for prompt electrons missed by the
FP TDC but not the FP scaler due to deadtime in the 
electronics. Typical values for $f_{\rm ghost}$ and $f_{\rm missed}$ were
5\% and 1\% respectively, while $f_{\rm stolen}$ ranged from 15 -- 50\% 
depending upon the FP rate for the runs in question. A summary of 
the values of the rate-dependent correction $f_{\rm R}$ is presented in 
Table~\ref{table:rate_and_target_corrs}, where the first uncertainty is a 
scale systematic uncertainty common to all data points and the second is a 
kinematic-dependent systematic uncertainty that arises uniquely from 
$f_{\rm stolen}$.

\subsubsection{\label{subsubsection:target_related_correction}Target-related Corrections}

The target-related correction was given by
\begin{equation}
  \label{equation:target_corrs}
  f_{\rm T} = f_{\rm abs} f_{\rm fill} f_{\rm cell},
\end{equation}

\noindent where $f_{\rm abs}$ was due to the absorption of beam photons by the liquid deuterium
prior to scattering, $f_{\rm fill}$ was due to incomplete filling of the target 
cell, and $f_{\rm cell}$ was due to beam photons scattering from the 
Kapton endcaps of the target cell.
The absorption of beam photons by the liquid deuterium prior to scattering was 
determined using a \geant\ Monte Carlo which considered the effective target 
thickness discussed in Sect~\ref{subsection:scale_normalization}. The correction
$f_{\rm abs}$ was determined to be $\sim$1.6\%, and was known to better than 
3\% relative uncertainty.

During RP1, the liquid deuterium target did not fill completely. 
The liquid-deuterium level was observed each day by taking Polaroid images of the 
target cell which clearly showed the filled portion of the cell. Based on these 
images, it was determined that the top $\sim$1.6~cm of the target cell was 
unfilled. In order to account for beam photons that passed through this unfilled 
portion of the target, a Monte Carlo simulation similar to the one used to determine 
$\kappa_{\rm eff}$ was employed. The fraction of beam photons that struck the 
filled portion of the target was determined, taking into account both the angular 
divergence of the beam photons and the uncertainty in the observed target-fill 
line. The correction $f_{\rm fill}$ was determined to be 7\%, and was known to
better than 2.5\% relative uncertainty.

The contribution of the thin Kapton endcaps to the scattered-photon yield was 
investigated using a $\sim$1~cm thick Kapton target. These Kapton data were
subjected to the analysis detailed in Section~\ref{subsection:yield_extraction}
to extract the thick Kapton target yield $Y_{\rm Kapton}$. The resulting 
correction to the scattered-photon yield due to the thin Kapton endcaps was 
given by
\begin{equation}
  \label{equation:kapton_correction}
  f_{\rm cell} = \frac{Y - xY_{\rm Kapton}}{Y},
\end{equation}
where $x$ was a factor used to scale the thick Kapton target yield to the thin 
Kapton endcap yield. This scaling factor depended on the relative Kapton 
thicknesses, numbers of incident photons, and target geometries which affected
detector acceptances. The correction factors for DIANA (150$^\circ$) and BUNI 
(120$^\circ$) were consistent with unity within uncertainties. The average 
correction factor or CATS (60$^\circ$) was 94\%. A systematic uncertainty of 2\% in
$f_{\rm cell}$ was determined for each detector. The target-related
corrections are given in Table~\ref{table:rate_and_target_corrs}.

\begin{table*}
\caption{\label{table:rate_and_target_corrs}
Rate- and target-dependent corrections. The first uncertainty in $f_{\rm R}$
is a systematic that is common to all data points, while the second uncertainty 
is a kinematic-dependent systematic. The uncertainties in $f_{\rm T}$ are systematic. The upper four
energy bins are from RP1 and the lower four from RP2.}
\begin{ruledtabular}
\begin{tabular}{r|rr|rr|rr}
$E_{\gamma}$ &               \multicolumn{2}{c|}{60$^\circ$} &             \multicolumn{2}{c|}{120$^\circ$} &               \multicolumn{2}{c}{150$^\circ$} \\
       (MeV) &                \multicolumn{1}{c}{$f_{\rm R}$} &      \multicolumn{1}{c|}{$f_{\rm T}$} &                \multicolumn{1}{c}{$f_{\rm R}$} &     \multicolumn{1}{c|}{$f_{\rm T}$} &                \multicolumn{1}{c}{$f_{\rm R}$} &      \multicolumn{1}{c}{$f_{\rm T}$} \\
\hline
        69.6 & 1.49 $\pm$ 0.05 $\pm$ 0.03 &  1.03 $\pm$ 0.03 & 1.60 $\pm$ 0.05 $\pm$ 0.02 & 1.09 $\pm$ 0.04 & 1.49 $\pm$ 0.05 $\pm$ 0.04 & 1.09 $\pm$ 0.04 \\
        77.9 & 1.40 $\pm$ 0.04 $\pm$ 0.03 &  1.03 $\pm$ 0.03 & 1.49 $\pm$ 0.05 $\pm$ 0.02 & 1.09 $\pm$ 0.04 & 1.42 $\pm$ 0.04 $\pm$ 0.03 & 1.09 $\pm$ 0.04 \\
        86.1 & 1.31 $\pm$ 0.04 $\pm$ 0.02 &  1.02 $\pm$ 0.03 & 1.36 $\pm$ 0.04 $\pm$ 0.01 & 1.09 $\pm$ 0.04 & 1.33 $\pm$ 0.04 $\pm$ 0.02 & 1.09 $\pm$ 0.04 \\
        93.4 & 1.28 $\pm$ 0.04 $\pm$ 0.02 &  1.02 $\pm$ 0.03 & 1.32 $\pm$ 0.04 $\pm$ 0.01 & 1.09 $\pm$ 0.04 & 1.30 $\pm$ 0.04 $\pm$ 0.02 & 1.09 $\pm$ 0.04 \\
\hline
        85.8 & 1.30 $\pm$ 0.04 $\pm$ 0.01 & 0.96  $\pm$ 0.02 & 1.58 $\pm$ 0.05 $\pm$ 0.02 & 1.02 $\pm$ 0.02 & 1.30 $\pm$ 0.04 $\pm$ 0.02 & 1.02 $\pm$ 0.02 \\
        94.8 & 1.26 $\pm$ 0.04 $\pm$ 0.01 & 0.96  $\pm$ 0.02 & 1.48 $\pm$ 0.05 $\pm$ 0.01 & 1.02 $\pm$ 0.02 & 1.26 $\pm$ 0.04 $\pm$ 0.02 & 1.02 $\pm$ 0.02 \\
       103.8 & 1.21 $\pm$ 0.04 $\pm$ 0.01 & 0.96  $\pm$ 0.02 & 1.39 $\pm$ 0.04 $\pm$ 0.01 & 1.02 $\pm$ 0.02 & 1.22 $\pm$ 0.04 $\pm$ 0.01 & 1.02 $\pm$ 0.02 \\
       112.1 & 1.19 $\pm$ 0.04 $\pm$ 0.01 & 0.96  $\pm$ 0.02 & 1.34 $\pm$ 0.04 $\pm$ 0.01 & 1.02 $\pm$ 0.02 & 1.20 $\pm$ 0.04 $\pm$ 0.01 & 1.02 $\pm$ 0.02 \\
\end{tabular}
\end{ruledtabular}
\end{table*}

\subsection{\label{subsection:uncertainities}Uncertainties}

The dominant contribution to the statistical uncertainty came from the yield. 
The systematic uncertainties were divided into three
classes: scale, angle-dependent, and kinematic-dependent. Scale uncertainties
affected the data obtained at all angles and energies in a given run period 
equally. Angle-dependent uncertainties affected the results from the individual
detectors differently. Kinematic-dependent uncertainties affected each
measured data point individually. We report these uncertainties separately.
The sources of systematic uncertainties are listed in Table~\ref{table:systs} 
along with typical values. 

\begin{table}
\caption{\label{table:systs}
Sources and magnitudes of systematic uncertainties for the two run periods.}
\begin{ruledtabular}
\begin{tabular}{llrr}
     Type &              Source &      RP1  &    RP2 \\
\hline
    Scale &  Tagging Efficiency &     1.1\% &   1.6\% \\
          &    Target Thickness &     2.5\% &   2.5\% \\
          &        Missed Trues &     1.5\% &   1.5\% \\
          &        Stolen Trues &     1.0\% &   1.0\% \\
          &        Ghost Events &     2.4\% &   2.4\% \\
          &         Kapton Cell &     2.0\% &   2.0\% \\
          &   Target Fill Level &     2.5\% &   N/A \\
\hline
    Total &                     &     5.2\% &   4.7\% \\
\hline\hline
\multicolumn{4}{c}{Other uncertainties}\\
\hline
          & Statistical         &    8--17\% &     5--12\% \\
\hline
  Angular & Detector Acceptance &     3--4\% &      3--4\% \\
\hline
Kinematic &    Yield Extraction &     2--11\% &     2--11\% \\
          &        Stolen Trues &     1--3\%  &      1--3\% \\
\end{tabular}
\end{ruledtabular}
\end{table}

\section{\label{section:results}Results}

\subsection{Cross Sections} \label{subsubsection:Cross Sections}

\begin{figure}
\begin{center}
\resizebox{\columnwidth}{!}{\includegraphics[trim = 0mm 0mm 10mm 0mm, clip]{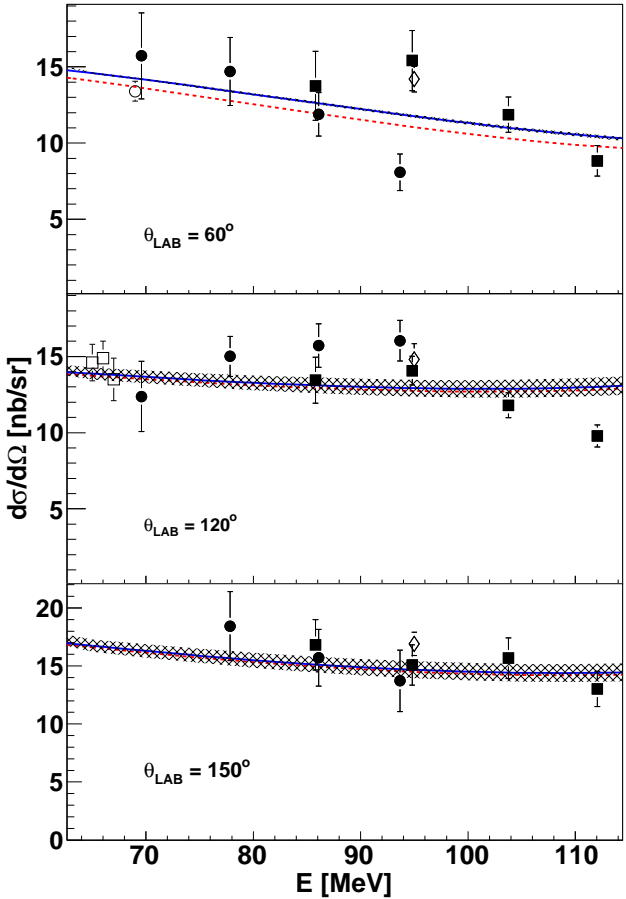}}
\caption{(Color online) Measurements of the deuteron Compton-scattering cross 
section from the current experiment (RP1 [$\bullet$] and RP2 [$\blacksquare$]) 
are shown. Previous measurements from References 
\cite{lucas1994} ($\circ$), \cite{lundin2003} ($\Box$), and \cite{hornidge2000} 
($\diamond$) are included for comparison. Statistical uncertainties only are shown.
The solid and dashed lines represent the free and BSR-constrained fits to the 
present data only. The shaded region is obtained by varying the BSR-constrained fit for $\alpha_s$ and 
$\beta_s$ by its statistical uncertainty.
\label{figure:figure_08_world_data_theory}}
\end{center}
\end{figure}

With the data analyzed as described above, we present the central result
of our experiment in Table~\ref{table:data}: elastic Compton scattering cross sections on the
deuteron. The results are also shown in
Fig.~\ref{figure:figure_08_world_data_theory}, along with those results at 66~MeV 
from Refs.~\cite{lucas1994,lundin2003} and 94~MeV from Ref.~\cite{hornidge2000} 
whose scattering angles are within $10^\circ$ of ours. Statistical uncertainties 
only are shown on the data points.

\begin{table*}
  \caption{\label{table:data}
    Measured cross sections for deuteron Compton scattering at the lab angles listed. 
    The first uncertainty is statistical, the second is the overall normalization 
    (5.2\% first run period, 4.7\% second), the
    third is due to detector acceptance, and the last is kinematic-dependent.
    The upper four energy bins are from RP1 and the lower four from RP2.}
\begin{ruledtabular}
\begin{tabular}{crrr}
$E_{\gamma}$ &   \multicolumn{1}{c}{$\frac{d\sigma}{d\Omega}(60^\circ)$} &   \multicolumn{1}{c}{$\frac{d\sigma}{d\Omega}(120^\circ)$} &        \multicolumn{1}{c}{$\frac{d\sigma}{d\Omega}(150^\circ)$} \\
       (MeV) &                               \multicolumn{1}{c}{(nb/sr)} &                               \multicolumn{1}{c}{(nb/sr)} &                                     \multicolumn{1}{c}{(nb/sr)} \\
\hline
        69.6 &   15.7  $\pm$ 2.6 $\pm$ 0.8 $\pm$ 0.7 $\pm$ 1.0                     &    12.4  $\pm$ 2.2 $\pm$ 0.6 $\pm$ 0.5 $\pm$ 0.8                      &                   \multicolumn{1}{c}{------------}\\
        77.9 &   14.7  $\pm$ 2.0 $\pm$ 0.8 $\pm$ 0.6 $\pm$ 0.9                     &    15.0  $\pm$ 1.3 $\pm$ 0.8 $\pm$ 0.6 $\pm$ 0.3                      &                   18.4  $\pm$ 2.5 $\pm$ 1.0 $\pm$ 0.6 $\pm$ 1.5 \\
        86.1 &   11.9  $\pm$ 1.4 $\pm$ 0.6 $\pm$ 0.5 $\pm$ 0.3                     &    15.7  $\pm$ 1.4 $\pm$ 0.8 $\pm$ 0.7 $\pm$ 0.3                      &                   15.7  $\pm$ 2.3 $\pm$ 0.8 $\pm$ 0.5 $\pm$ 0.8 \\
        93.4 &    8.1  $\pm$ 1.2 $\pm$ 0.4 $\pm$ 0.3 $\pm$ 0.3                     &    16.0  $\pm$ 1.3 $\pm$ 0.8 $\pm$ 0.7 $\pm$ 0.4                      &                   13.7  $\pm$ 2.2 $\pm$ 0.7 $\pm$ 0.4 $\pm$ 1.5 \\
\hline
        85.8 &   13.8  $\pm$ 1.7 $\pm$ 0.6 $\pm$ 0.6 $\pm$ 1.5                     &    13.4  $\pm$ 1.0 $\pm$ 0.6 $\pm$ 0.6 $\pm$ 1.2                      &                   16.8  $\pm$ 2.0 $\pm$ 0.8 $\pm$ 0.5 $\pm$ 0.7 \\
        94.8 &   15.4  $\pm$ 1.5 $\pm$ 0.7 $\pm$ 0.7 $\pm$ 1.2                     &    14.1  $\pm$ 0.8 $\pm$ 0.7 $\pm$ 0.6 $\pm$ 0.5                      &                   15.1  $\pm$ 1.7 $\pm$ 0.7 $\pm$ 0.5 $\pm$ 0.5 \\
       103.8 &   11.9  $\pm$ 1.1 $\pm$ 0.6 $\pm$ 0.5 $\pm$ 0.3                     &    11.8  $\pm$ 0.7 $\pm$ 0.6 $\pm$ 0.5 $\pm$ 0.4                      &                   15.7  $\pm$ 1.6 $\pm$ 0.7 $\pm$ 0.5 $\pm$ 0.8 \\
       112.1 &    8.8  $\pm$ 1.0 $\pm$ 0.4 $\pm$ 0.4 $\pm$ 0.2                     &     9.8  $\pm$ 0.7 $\pm$ 0.5 $\pm$ 0.4 $\pm$ 0.2                      &                   13.0  $\pm$ 1.5 $\pm$ 0.6 $\pm$ 0.4 $\pm$ 0.4 \\
\end{tabular}
\end{ruledtabular}
\end{table*}

The current world data set of deuteron Compton scattering cross sections is
comprised of three measurements \cite{hornidge2000,lucas1994,lundin2003}.  The
data reported here have uncertainties comparable to the previous data at low
energies (E~$\le$~70~MeV) and energy-bin widths considerably smaller than the
previous high-energy (E~=~95~MeV) measurement (Table~\ref{table:world_comp}).
This experiment has doubled the number of data points in the world data set, in
addition to providing the first data above 100~MeV. As
Fig.~\ref{figure:figure_08_world_data_theory} indicates, all data sets are in
excellent agreement within their respective statistical uncertainties. This is
corroborated by a complementing angle transect at $66$ and $94.5$~MeV, where
these preceding measurements provide additional data beyond our scattering angles; see
Ref.~\cite{myers2014a} for plots. Data consistency in the overlap is already a 
strong indication that our results can well be embedded into the consistent 
world data set, extending it to higher energies.

\begin{table}
\caption{\label{table:world_comp}
Comparison of reported results from measurements of deuteron Compton 
scattering.}
\begin{ruledtabular}
\begin{tabular}{lcccc}
  Ref. & E & $\Delta$E & Number & Normalization \\
  & (MeV) & (MeV) & of Points & Uncertainty \\
  \hline
  \ \cite{lucas1994} & 49,69 & 6.5,7.7 & 6 & 4\% \\
  \ \cite{hornidge2000} & 95 & 21 & 5 & 5--6\% \\
  \ \cite{lundin2003} & 55,67 & 10 & 18 & 6--14\% \\
  This work & 70-112 & 7.3-9.0 & 23 & 5\% \\ 
\end{tabular}
\end{ruledtabular}
\end{table}

The figures also point to issues with two of our new data. The two points around
$94.5$~MeV and $60^\circ$ are separated by $\sim$3$\sigma$, with the one from RP2 
well in agreement with a SAL measurement and better agreeing with
the overall trend. Another, more subtle, compatibility issue with the $(112.1$~MeV$,120^\circ)$
point of the second run arises when comparing the data to fits, see also Refs.~\cite{myers2014a,futuretheory}. For both
instances, the discrepancies could not be traced back to specific data or
analysis issues. We therefore report these points but flag them as potential
outliers, presenting results with them included in our extraction of the
isoscalar nucleon polarizabilities in this paper. 

\subsection{Nucleon Polarizabilities} \label{subsubsection:Polarizabilities}

As argued in the Introduction, the goal of this experiment was to provide new,
high-quality data in support of an extraction of the neutron
polarizabilities. Such a determination of these two-photon response parameters
needs theoretical input, but it will also allow for a better understanding of
the quality of our data.

First, the effects of nuclear binding and charged meson-exchange currents
inside the deuteron must be subtracted. They contribute a significant fraction
of the deuteron cross section at the energies we measured~\cite{Beane99}, as
well as in the zero-energy limit in which $\alphae$ and $\betam$ are
defined~\cite{Hildebrandt05}. For the individual amplitudes, one then needs to
subtract the Powell amplitudes of point-like spin-$\nicefrac{1}{2}$ particles with
anomalous magnetic moments.
 
More importantly, however, the nonzero-energy data must be related to the
static point. In a simplistic extrapolation, one may be tempted to use a 
cross-section fit which identifies the electric and magnetic polarizabilities as
angle-dependent contributions to the terms quadratic in the photon
energy. This is, however, not permissible since all of our new data, and most of
the world data, are well beyond its realm of applicability. At energies at and
above $100$~MeV, the energy-dependent effects of the pion cloud and of the
$\Delta(1232)$ excitation are important, and the pion-production threshold
induces the first non-analyticity in the single-nucleon amplitudes. A
low-energy expansion proceeds thus in powers of $\omega/m_\pi$ and becomes
quickly useless.

A viable low-energy parametrization should hence consistently account for all
these effects and provide a systematically improvable estimate of residual
uncertainties. Chiral Effective Field Theory (\ChiEFT) is ideally suited for
this task. It model-independently encodes the correct symmetries and effective
degrees of freedom of low-energy QCD, with a small dimensionless parameter to
systematically improve the description of higher-order effects. It is well
established that \ChiEFT predicts the energy dependence of the single-nucleon
Compton scattering response over the full range of data with high accuracy,
including spin effects~\cite{griesshammer2012}, and that it consistently accounts for nuclear binding
as well as meson-exchange and nucleon-nucleon rescattering effects. All these
aspects are necessary to restore the Thomson limit on the deuteron. 

We therefore turn to the \ChiEFT description which was used in previous
high-accuracy descriptions of proton and deuteron Compton scattering. As its
ingredients at next-to-leading order in $\alphae$ and $\betam$ (order
$e^2\delta^3$) have recently been described summarily, we refer to Sect.~5.3
of Ref.~\cite{griesshammer2012} for details. The interactions between
nucleons, pions and the $\Delta(1232)$ resonance are fully determined except for
the two which parametrize short-distance contributions to the scalar
polarizabilities. The dependence of the resulting extraction of $\alpha$ and
$\beta$ on the deuteron wave function and $NN$ potential was shown to be
negligible, and residual theoretical uncertainties were estimated as $\pm0.8$
canonical units. In the results reported here, we use the same fit procedure and
parameters as reported in Ref.~\cite{griesshammer2012}. As described there, we add kinematic-dependent and angle-dependent
systematic uncertainties in quadrature to the statistical uncertainty, and treat
correlated overall systematic uncertainties by a floating normalization. This
determination is based on the entire data set of our experiment, but does not
include the other world data. Finally, we treat the two run periods as
statistically independent data sets. Treating them as a single data set changes the following
conclusions at most marginally.

Table~\ref{table:pols} summarizes the findings of our fit, in the context of the
previous extraction of the polarizabilities and the determination based on the 
new world data set which includes our data but with the two points previously 
mentioned discarded -- these results being already
reported in Ref.~\cite{myers2014a} and Eq.~\eqref{eq:AlphaBetaNewBaldin}. For
our extraction, as for the others, the values of the independent fit to
$\alpha_s$ and $\beta_s$ demonstrate excellent consistency of $\alpha_s+\beta_s$
with the isoscalar BSR, Eq.~\eqref{eq:BSRS}. The BSR can therefore
be used to reduce the number of parameters from two to one, decreasing the
statistical uncertainties.  In the fit to only the new data, the normalization
of RP1 floats by about $2$\% down, and that of RP2
by $3.5$\% up, i.e.~well within the overall correlated uncertainty. The statistical
uncertainties are smaller than for the previous world data set, but the
$\chi^2$ per degree of freedom is larger. As hinted above, this can be
attributed to two points which between themselves contribute about $20$ units
to $\chi^2$, while changing the central values only within the statistical
uncertainties. A complementary publication~\cite{futuretheory} provides details in
the context of the construction of a consistent database. Here, we reiterate
that a careful analysis of our data-taking and analysis procedures showed no
intrinsic experimental reason why these points should be special.

\begin{table}
  \caption{\label{table:pols}
    Isoscalar nucleon polarizabilities and $\chi^2$ per degree of freedom extracted from the world elastic deuteron
    Compton scattering data set prior to our data, from our data, and from the
    new world data set including our data, each using both a ``free'' fit to
    $\alpha_s$ and $\beta_s$ and a determination constrained by the BSR. 
    Uncertainties are statistical only and anticorrelated for the
    BSR-constrained fits.}
\begin{ruledtabular}
\begin{tabular}{lrrrrr}
set & constraint & $\alpha_s$& $\beta_s$& $\chi^2$/d.o.f & Ref.\\
\hline
\multirow{2}{*}{old world} 
& free & $10.5\pm2.0$& $3.6\pm1.0$&$24.3/24$&\\ 
&BSR &$10.9\pm0.9$&$3.6\mp0.9$&$24.3/25$&
        \multirow{2}{*}{\cite{griesshammer2012}}\\\hline\hline
\multirow{2}{*}{this work}
& free & $13.2\pm1.4$& $3.2\pm1.1$&$40.6/19$& 
       \\
&BSR &$12.1\pm0.8$&$2.4\mp0.8$&$41.7/20$&\\\hline
\hline
\multirow{2}{*}{new world} 
& free & $11.1\pm0.9$& $3.3\pm0.6$&$49.2/43$& \\
&BSR &$11.1\pm0.6$&$3.4\mp0.6$&$45.2/44$&
        \multirow{2}{*}{\cite{myers2014a}}\\
\end{tabular}
\end{ruledtabular}
\end{table}

Our central values for either fit agree with those of the old and new world
database extractions well within the systematic uncertainties only, not accounting for
theoretical and BSR uncertainties. Since it effectively doubles the world data, it is therefore no surprise that the new
world average reported in Ref.~\cite{myers2014a} is hardly shifted but its
statistical uncertainty is reduced to $1/\sqrt{2}\approx70\%$ of the previous
one.

With only our current data, we thus obtain the final values of the BSR-constrained
fit as 
\begin{equation}
  \label{eq:AlphaBetaSMyersOnly}
  \begin{split}
    &\alpha_s = 12.1 \pm 0.8(\text{stat}) \pm 0.2(\text{BSR}) \pm
    0.8(\text{th}) \\ 
    &\beta_s = 2.4 \mp 0.8(\text{stat}) \pm 0.2(\text{BSR}) \mp
    0.8(\text{th}),
  \end{split}
\end{equation}
where the uncertainties from theory and from the BSR constraint
are listed separately. One finally extracts the neutron polarizabilities by
combining with the proton values of Eq.~\eqref{eq:AlphaBetaP} which were
determined using the same \ChiEFT approach and fit philosophy
\begin{equation}
  \label{eq:AlphaBetaNMyersOnly}
  \begin{split}
    &\alpha_n = 13.55 \pm 1.6(\text{stat}) \pm 0.2(\text{BSR}) \pm
    0.8(\text{th}) \\ 
    &\beta_n = 1.65 \mp 1.6(\text{stat}) \pm 0.2(\text{BSR}) \mp 0.8(\text{th}).
  \end{split}
\end{equation}
We list these values for completeness only, since those obtained from the new,
statistically consistent world data set~\cite{myers2014a} supersede them 
in accuracy
\begin{equation}
  \label{eq:AlphaBetaNnew}
  \begin{split}
    &\alpha_n = 11.55 \pm 1.25(\text{stat}) \pm 0.2(\text{BSR}) \pm
    0.8(\text{th}) \\ 
    &\beta_n = 3.65\mp 1.25(\text{stat}) \pm 0.2(\text{BSR}) \mp 0.8(\text{th}).
  \end{split}
\end{equation}

In conclusion, careful statistical tests indicate that our measurement
provides new, high-quality data whose analysis is well-understood.

\section{\label{section:summary}Summary}

This paper reports new Compton-scattering 
cross sections for deuterium over an energy range from 70--112 MeV, 
at angles of $60^\circ$, $120^\circ$ and $150^\circ$.
The data points are in excellent agreement with previously
published results. An analysis using \ChiEFT\ extracts isoscalar 
nucleon polarizabilities that agree, within uncertainties, with
previous extractions. This measurement represents the first new result 
from deuteron Compton scattering in more than ten years, nearly 
doubles the number of data points in the global data set, and 
extends the maximum energy by almost 20~MeV. Furthermore, this data 
set reduces the statistical uncertainty of the extracted values of 
$\alpha_n$ and $\beta_n$ by a factor of $\sim$$1/3$.

\begin{acknowledgments}

The authors acknowledge the support of the staff of the MAX IV Laboratory. We 
also gratefully acknowledge the Data Management and Software Centre, a Danish 
contribution to the European Spallation Source ESS AB, for generously providing 
access to their computations cluster. We would like to thank J.~A. McGovern 
and D.~R. Phillips for discussions about the extraction of the polarizabilities;
H.W.G. thanks both for their continuing collaboration on the theoretical aspects.
The Lund group acknowledges the financial 
support of the Swedish Research Council, the Knut and Alice Wallenberg Foundation, 
the Crafoord Foundation, the Wenner-Gren Foundation, and 
the Royal Swedish Academy of Sciences. This work was sponsored in part by the 
 US National Science Foundation under Award Number 0855569, 
the U.S. Department of Energy under grants 
DE-FG02-95ER40907 
and DE-FG02-06ER41422, 
and the UK Science and Technology Facilities Council under grants
57071/1 and 50727/1. 

\end{acknowledgments}

\bibliography{myers_etal}

\end{document}